\documentclass[letter]{aa}
\usepackage{graphicx}
\usepackage{amssymb}

\usepackage{txfonts}
\usepackage{color}
\usepackage{natbib}
\usepackage{tabularx}
\usepackage{multirow}

\begin{document}

\title{Occurrence of potentially hazardous GRBs launched in globular clusters}

\author{Wilfried F. Domainko\inst{1}}

\institute{Max-Planck-Institut f\"ur Kernphysik, P.O. Box 103980, D 69029
Heidelberg, Germany}

\offprints{\email{wilfried.domainko@mpi-hd.mpg.de}}

\date{}
 
\abstract
{Nearby, Galactic gamma-ray bursts (GRBs) may affect the terrestrial biota 
if their radiation is beamed
towards the Earth. Compact stellar binary mergers are possible central engines
of short GRBs and their rate could be boosted in globular clusters.}
{The occurrence, impact on the terrestrial biota and the point in time of the explosion
of the closest GRB launched in a globular cluster are explored.}
{Globular cluster typically follow well defined orbits around the galactic center. 
Therefore their position relative to the solar system can be calculated back in time.
This fact is used to demonstrate that globular cluster - solar system encounters 
define possible points in time 
when a nearby GRB could have exploded.
Additionally, potential terrestrial signatures in the geological record connected to 
such an event are discussed.}
{Assuming rates of GRBs launched in globular cluster found from the
redshift distribution of short burst and adopting the current
globular cluster space-density around the solar system
it is found that the expected minimal distance $d_\mathrm{min}$
for such a GRB in the last Gyr is in the range 
$d_\mathrm{min} \sim 1 - 3.5$~kpc. From the average gamma-ray
luminosity of a short GRB significant depletion of the terrestrial ozone-layer is 
expected if such an event explodes at a distance of $\sim1$~kpc. In the last Gyr
a few globular cluster passages are expected within a distance of $d_\mathrm{min}$
from the solar system and a GRB should have exploded during one of these passages.}
{Globular cluster - solar system encounters and events of mass 
extinction in the history of life can be correlated to investigate the impact
of a nearby GRB on the terrestrial biota. To explore such a correlation 
reliable globular cluster positions relative to the solar system have to be 
calculated for the time span of the fossil record of the last 600~Myrs. 
The upcoming GAIA mission will
be crucial to determine the possible time intervals of the occurrence of nearby GRBs 
launched in globular clusters.}
\keywords{globular clusters: general -- Gamma-ray burst: general -- Proper motions -- Astrobiology}

\authorrunning{W. Domainko}
\titlerunning{Potentially hazardous GRBs in globular clusters}

\maketitle


\section{Introduction}

Gamma-ray bursts (GRBs) are powerful cosmological explosions which 
irradiate their surroundings in two cones with high energy photons 
\citep[see][for a review]{gehrels2009}. It has been recognized that
due to this radiation, galactic GRBs can have a dramatic effect on 
the earth atmosphere and on the biota \citep{thorsett1995}.
Furthermore, it has been pointed out that beside the extinction of species
cosmic rays connected to GRBs can cause biological mutations
leading to fast appearance of new species after these mass extinctions events
\citep{dar1998}.
Indeed the evolution of live on earth
was affected by several events of mass-extinction \citep{raup1982}
and show also periods of rapid development of new species 
\citep[e.g. Cambrian explosion][]{marshall2006}.
From the observed rate of GRBs it is evident that a nearby ($d \sim 1$~kpc)
burst is likely during the geological record and such an event
has been linked to the late Ordovician mass extinction event 
that happened 440~Myr in the past \citep{melott2004}.
Additionally,  \citet{horvath2003} has
suggested that the Cambrian explosion 543~Myr ago could have 
been triggered by a nearby GRB.
In general, the exact point in time when a nearby GRB has happened
is difficult to identify. This challenges the correlation between
a certain event in the history of the Earth and the occurrence of a nearby GRB.

GRBs appear to be launched by two distinct populations of progenitors.
Long bursts (duration $> 2$~s) appear to be connected to the death of
massive stars whereas short burst (duration~$<~2$~s) seem to result from
the merger of two compact objects \citep[see][for a review]{gehrels2009}.  

For short bursts it has been shown that their rate can be boosted in
globular clusters due to the efficient production of
close stellar binaries in their cores \citep{grindlay2006,lee2010}.
Several observational facts support the association between short
GRBs and globular clusters.  
Firstly, in the globular cluster
M~15 a close binary consisting of two neutron stars has already 
been discovered \citep{anderson1990}. Secondly, from the off-set distribution
of short bursts from their host galaxies it has been inferred that
a subset of bursts can originate in globular clusters 
\citep{berger2010,salvaterra2010,church2011}.
Thirdly, from the redshift distribution of short burst it has been concluded that
in the local universe the short GRB rate is dominated by dynamically 
formed compact binaries in globular clusters \citep{salvaterra2008,guetta2009}.
Finally the occurrence of short GRBs in globular clusters in the Milky Way
can also be probed by the presence of corresponding remnants 
\citep{domainko2005,domainko2008}. Recently,
indeed a source of very-high-energy gamma-rays, HESS~J1747-248, 
in the direction of the
globular cluster Terzan~5 has been discovered \citep{abramowski2011}. This
source is potentially accompanied by diffuse X-ray emission \citep{eger2010}
and diffuse radio emission \citep{clapson2011} and has been interpreted in the 
framework of a GRB remnant \citep{domainko2011}. To conclude there 
seems to be
evidence that a considerable rate of GRBs are launched in globular clusters.

Globular clusters typically follow well defined orbits around
the center of the Milky Way.
Proper motions are available for a number of globular clusters
\citep{dinescu1997,dinescu1999a,dinescu1999b} and this allows
to determine their orbital motions \citep{allen2006}. 
Thus, their position relative to
the Earth can be traced back in time \citep{vandeputte2009}. Encounters
of globular clusters and the Earth are potential time intervals during
which to the biota hazardous, nearby bursts can occur.
Consequently it is possible to identify such periods by tracing back 
the distance of globular cluster to the Earth.

The observed values for the proper motion of globular clusters
are still affected by considerable errors. \citet{vandeputte2009}
found that the relative position between the Earth and 
various globular
clusters can be traced back in time reliably only for about 50~Myr.
Since this time span is considerably shorter than the geological 
record, in this paper the possibility of a correlation between 
globular cluster passages
and events in the history of the Earth are investigated on a
statistical basis. 

This paper is organized as follows: firstly the expected minimum distance
to a GRB launched in a globular cluster will be determined. Secondly, potential
terrestrial signatures connected to a GRB happening at this minimal distance
are investigated. Thirdly, it will be shown that
with determining the time intervals of globular cluster - solar system encounters  
points in time can be identified, where in principal nearby GRBs can happen.
And finally the prospects for correlating such globular cluster passages and
events of mass extinction are discussed.


\section{Minimal distance}
\label{sec:mindist}

To determine the expected minimum distance to Earth of a GRB launched in 
a globular cluster in the last Gyr, first the rate of such events per 
globular cluster has to be constrained. The estimated rate of GRBs originating in
globular clusters that are beamed towards the Earth 
ranges from $R_{GC} \sim 4$\,Gpc$^{-3}$yr$^{-1}$ \citep{guetta2009} to
$R_{GC} \sim 20$\,--\,90\,Gpc$^{-3}$yr$^{-1}$ \citep{salvaterra2008}.  
Using now the density of globular clusters
in the local universe\footnote{assuming $h$= 0.7} of about 3~Mpc$^{-3}$ 
\citep{portegies2000} this results in a GRB rate of $\Phi \approx 10^{-8}\,(R_{GC}/30\, 
\mathrm{Gpc}^{-3}\mathrm{yr}^{-1})$ per year per globular cluster. From this estimate
the expected minimum distance $d_{min}$ for a GRB exploding in the last Gyr can be
estimated. For an event rate of 
10$^{-9}$ per year $d_{min}$ is given by the 
volume around the Earth where the probability
to find a globular cluster is $10^{-9}/\Phi \approx 0.1 \times (R_{GC}/30\, 
\mathrm{Gpc}^{-3}\mathrm{yr}^{-1})^{-1}$. The resulting minimal distance is then
given by:

\begin{equation}
d_{min} \approx \left[ \frac{3}{4\, \pi} \left( \frac{0.1 \times R_{GC}}{30\, \mathrm{Gpc}^{-3}\mathrm{yr}^{-1}}\right)^{-1} \left (\frac{\rho_{GC}}{\mathrm{kpc}^{-3}}\right)^{-1} \right]^{1/3}~\mathrm{kpc}
\end{equation}

\begin{figure}[ht]
\centering
\includegraphics[height=6cm]{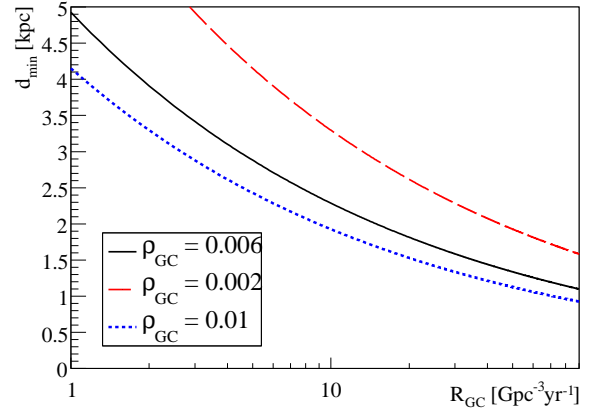}
\caption{Dependence of the minimal distance to a GRB launched in a globular cluster 
in the last Gyr on the parameters GRB rate ($R_\mathrm{GC}$) and Galactic globular
cluster density at the position of the Earth ($\rho_\mathrm{GC}$). Three
curves are shown: $\rho_\mathrm{GC} = 0.01~$kpc$^{-3}$, 
$\rho_\mathrm{GC} = 0.006$~kpc$^{-3}$ (the present
day value) and $\rho_\mathrm{GC} = 0.002$~kpc$^{-3}$.}
\label{figure:dmin}
\end{figure}

Here $\rho_\mathrm{GC}$ is the number density of globular clusters around the Earth.
Adopting a local number density of
globular clusters of 0.006~kpc$^{-3}$ 
\citep{djorgovski1994} this results in a minimum
distance of about 1.6~kpc for a short GRB rate launched in globular clusters
for $R_{GC} = 30\, \mathrm{Gpc}^{-3}\mathrm{yr}^{-1}$. The dependence of
$d_{min}$ on the parameters $R_\mathrm{GC}$ and $\rho_\mathrm{GC}$ is
shown in Fig. \ref{figure:dmin}. For reasonable input parameters $d_\mathrm{min}$
falls between $\sim1 - 3.5$~kpc.

\section{Terrestrial signatures}

Potential terrestrial signatures generated by a nearby GRB could comprise
the impact on the biota in-printed in the fossil record 
\citep[see][for a discussion on various to the biota 
hazardous effects connected to a nearby burst]{galante2007},
a glaciation event following reduced atmospheric transparency due to
NO$_2$ production by the ionizing radiation of a GRB \citep{thomas2005}, 
anomalies of radioactive isotopes or fossil cosmic ray tracks \citep{dar1998}. 
From the expected minimal distance of a GRB launched in a globular cluster 
the strength of corresponding terrestrial signatures can be estimated.

\subsection{Impact on the biota}

Depletion of the ozone-layer has been found to be an important
consequence of a nearby GRB explosion for the biota \citep{thorsett1995,thomas2005}. 
The level of depletion of the ozone-layer will depend on the fluence
of X-rays and soft gamma-rays from the GRB.
If the mean isotropic gamma-ray luminosity 
$E_\mathrm{\gamma,iso} \approx 6.5 \times 10^{50}$~erg found for short
bursts by \emph{Swift}-BAT \citep{racusin2011} is adopted this results in a fluence
of $6 \times 10^{6}$~erg~cm$^{-2}$ at a distance of 1~kpc. 
In comparison to that value, \citet{galante2007} estimated the critical 
fluence of a GRB
to destroy the ozone layer on a level where damages on the biota will occur
to $3 \times 10^{6}$~erg~cm$^{-2}$ and also \cite{thomas2005}
found significant ozone depletion for the case of $10^7$~erg~cm$^{-2}$.
Consequently its possible that a short GRB at a distance of up to 1.5~kpc will have
an effect on the biota. To conclude, for advantageous estimates for the 
GRB rate in globular clusters it appears possible that the closest
such event has left terrestrial signatures in form
of an extinction event that is
in-printed in the fossil record.

\subsection{Radioactive nuclei}

One example 
for potentially measurable signatures are radioactive nuclei produced by either the 
enhanced cosmic ray flux \citep{dar1998} or by high energy gamma-ray photons 
\citep{thorsett1995} connected to the GRB.

Ultra-relativistic shock waves are expected to convert most of their kinetic 
energy into cosmic rays \citep{atoyan2006}. Following this argument
by assuming that the mean kinetic energy 
estimated for short bursts with \emph{Swift}-BAT of $2.5 \times 10^{52}$~erg 
\citep{racusin2011} is converted into cosmic rays and that the cosmic ray jet 
is still collimated at a distance of 1~kpc this will result in a cosmic ray 
fluence of $2 \times 10^8$~erg~cm$^{-2}$ at this distance. This corresponds to the integrated 
energy deposition of galactic cosmic rays for about 100~years.
Thus cosmic rays produced by the GRB can in principle create 
an anomaly of radioactive nuclei.
However, since terrestrial rocks are typically exposed to cosmic rays
much longer than 100~years its quite unlikely that cosmic rays connected to a GRB
will leave measurable traces in form of radioactive isotopes in the
geological record.

Some short GRBs show gamma-ray emission with photon energies above 100~MeV 
\citep{abdo2009} but it has to be noted that only few GRBs are detected
in this energy band \citep[e.g.][]{zhang2011}. 
For the case of GRB090510 the high energy component had an isotropic 
luminosity of about $4 \times 10^{52}$~erg \citep{abdo2009}. 
This high-energy luminosity can result in a comparable 
fluence of high-energy gamma-ray
photons as has been estimated for the cosmic ray fluence in the previous
paragraph. Consequently,
also the energy deposition of high energy gamma-ray photons can equal the 
galactic cosmic ray energy deposition integrated over about 
100~years if such a burst is located at a distance of 1~kpc.
However, again its quite unlikely that high-energy photons will 
leave measurable traces in the geological record in form
of radioactive nuclei.

To conclude, even for advantageous estimates for the GRB rate in globular cluster
its quite unlikely that these GRBs have left anomalies of 
radioactive isotopes in the geological record. However, for a short period of time 
($\lesssim 100$~years) they could have elevated the level of radioactivity
which could cause biological mutations
leading to fast appearance of new species.


\section{Association with terrestrial signatures}
\label{sec:asso}

Nearby GRBs launched in globular clusters offer the exciting possibility
to calculate back the approximate time of their occurrence. This can be done
due to the fact that globular clusters follow quite well defined orbits around
the galactic center \citep{allen2006} and therefore the time of encounters 
between the solar system and a specific globular cluster can be 
determined \citep[see e.g.][]{vandeputte2009}. Here a globular cluster
encounter is defined as a time period when the distance between the
solar system and the globular cluster is smaller than $d_\mathrm{min}$.

With the knowledge of the fraction of
time where at least one globular cluster can be found within $d_\mathrm{min}$
(see Sec. \ref{sec:mindist}) and with adopting the duration of a typical 
globular cluster -- Earth encounter the number of such encounters can be
estimated for a specific time interval. In Sec. \ref{sec:mindist} it was estimated that during
a fraction of about $0.1 \, (R_{GC}/30\, \mathrm{Gpc}^{-3}\mathrm{yr}^{-1})$ of
the time considered, one globular cluster is found within the minimal distance 
$d_\mathrm{min}$ for a GRB occurrence. A typical encounter will last
for $\mathcal{O}(10^7)$~years if an encounter length of $\sim$1~kpc and a relative
velocity between the earth and the globular cluster of $\sim$100~km\,s$^{-1}$
is assumed. If now a GRB rate
$R_{GC} = 30\, \mathrm{Gpc}^{-3}\mathrm{yr}^{-1}$ is considered this
results in a total time of $10^8$~years of presence of at least one globular cluster
within $d_\mathrm{min}$ in the last Gyr. With a typical encounter duration of
$\sim10^7$~years $\mathcal{O}(10)$
globular cluster -- Earth encounters within a distance of $d_\mathrm{min}$
are expected.

The expected probability for a GRB occurring in a
globular cluster during an encounter 
can be estimated from the rate of GRBs per globular cluster
and the duration of an globular cluster -- Earth encounter.
If for this rate a value of $\Phi \approx 10^{-8}\,(R_{GC}/30\, 
\mathrm{Gpc}^{-3}\mathrm{yr}^{-1})$ per globular cluster
per year (see Sec. \ref{sec:mindist})
is adopted and a typical duration of a globular cluster passage
of $10^7$~years is assumed, then the resulting probability is  
$\Phi \times 10^7\, \mathrm{years} \approx 0.1\,(R_{GC}/30\, 
\mathrm{Gpc}^{-3}\mathrm{yr}^{-1})$. For a GRB rate of
$R_{GC} = 30\, \mathrm{Gpc}^{-3}\mathrm{yr}^{-1}$ on average one
GRB is expected every 10 globular cluster -- Earth encounters.
Thus for the time span of the geological record one GRB is likely
during a globular cluster - solar system encounter.

In the above considerations it is assumed that the GRB rate
is the same for each globular cluster. However, some globular
clusters are especially prolific producers of close binaries \citep{pooley2006}.
These objects likely exhibit also a higher GRB rate \citep{grindlay2006,lee2010}.
Consequently, the passage of a globular cluster hosting a 
large number of stellar binaries close to Earth can be identified
as a potential time interval of a nearby GRB explosion.

As a result of the above considerations, specific time intervals corresponding to
globular cluster encounters can be identified for which the geological record
can be searched for potential terrestrial signatures connected to a nearby GRB.
Furthermore correlations between events of mass extinction in the history of
life \citep{raup1982} or periods of rapid development of new species 
\citep[e.g. Cambrian explosion][]{marshall2006} with periods of 
globular cluster encounters can be explored.
With a few globular cluster passages during the last 600~Myr and 
a typical duration of $\mathcal{O}(10^7)$~years for each passage
a fair fraction of chance coincidence between events of mass extinctions and
globular cluster passages can be expected. Considering only passages of globular
clusters hosting an elevated number of close binaries can help to reduce the
probability of such a chance coincidence.


\section{Outlook}

In Sec. \ref{sec:asso} it has been pointed out that periods
with globular cluster passages close to the solar systems can be determined.
During these time periods nearby GRB events launched in globular clusters
could have happened. For determining the impact of nearby GRBs 
on the biota by correlating
globular cluster passages and events of mass extinction, it is 
necessary to reliably calculate the globular cluster distance for the last
600~Myrs. An accurate determination of globular cluster positions appear in principle
possible since the time span of 600~Myrs corresponds to only a few
globular cluster orbits around the galactic center that last typically
$\mathcal{O}(100)$~Myrs.

Assuming a velocity of the globular clusters 
of $\mathcal{O}$(100)~km\,s$^{-1}$
an accuracy of about 1\% for the 3d motion
would be needed to constrain their position to 
$\sim1$~kpc for the last Gyr.
However, with the current precision reliable calculations
of the cluster position with respect to the Earth can only be
calculated back for about 50 Myrs ago \citep[see][]{vandeputte2009}.

The required accuracy for the 3d motion of galactic globular clusters
will likely be available in the foreseen future.
The up-coming GAIA mission\footnote{http://gaia.esa.int/} 
will provide unprecedented positional and radial velocity measurement 
of about one billion stars in the Milky Way. With these data it will
be possible to calculate accurate globular cluster orbits and to identify
periods of encounters with the solar system. Encounters between 
globular clusters with an exceptionally elevated number of close binaries
\citep{pooley2006} can be identified as potential point in time for a nearby
GRB explosion. These points in time can be compared to events of mass extinction
and for the case that there is no correlation with such events, the geological 
record can be searched for any terrestrial
signatures connected to a nearby GRB explosion.
 

\begin{acknowledgements}

The author acknowledge support from his host institution. The author want to thank
A.-C. Clapson and M. Ruffert for many enlightening discussions.

\end{acknowledgements}



\begin{thebibliography}{}

\bibitem[Abdo et al.(2009)]{abdo2009} Abdo, A. A. et al. (\emph{Fermi}-GBM/LAT collaboration) 2009, Nature, 462, 331

\bibitem[Abramowski et al.(2011)]{abramowski2011} Abramowski, A. et al. (H.E.S.S. collaboration) 2011, A\&A, 531, L18

\bibitem[Allen et al.(2006)]{allen2006} Allen, C., Moreno, E. \& Pichardo, B. 2006, ApJ, 652, 1150

\bibitem[Anderson et al.(1990)]{anderson1990} Anderson, S. B., Gorham, P. W., Kulkarni, S. R., Prince, T. A., \& Wolszczan, A. 1990, Nature 346, 42

\bibitem[Atoyan et al.(2006)]{atoyan2006} Atoyan, A., Buckley, J. \& Krawczynski, H. 2006, ApJ, 642, L153

\bibitem[Berger(2010)]{berger2010} Berger, E. 2010, ApJ, 722, 1946

\bibitem[Church et al.(2011)]{church2011} Church, R. P., Levan, A. J., Davies, M. B., \& Tanvir, N. 2011, MNRAS, 413, 2004

\bibitem[Clapson et al.(2011)]{clapson2011} Clapson, A.-C., Domainko, W., Jamrozy, M., Dyrda, M. \& Eger, P. 2011, A\&A, 532, 47

\bibitem[Dar et al.(1998)]{dar1998} Dar, A., Laor, A. \& Shaviv, N. J. 1998, PRL, 80, 5813

\bibitem[Dinescu et al.(1997)]{dinescu1997} Dinescu, D. I., Girard, T. M., van Altena, W. F., Mendez, R. A. \& L\'opez, C. E. 1997, AJ, 114, 1014

\bibitem[Dinescu et al.(1999a)]{dinescu1999a} Dinescu, D. I., van Altena, W. F., Girard, T. M. \& L\'opez, C. E. 1999a, AJ, 117, 277

\bibitem[Dinescu et al.(1999b)]{dinescu1999b} Dinescu, D. I., Girard, T. M., van Altena, W. F. 1999b, AJ, 117, 1792

\bibitem[Djorgovski \& Meylan(1994)]{djorgovski1994} Djorgovski, S. \& Meylan, G. 1994, AJ, 108, 1292

\bibitem[Domainko \& Ruffert(2005)]{domainko2005} Domainko, W. \& Ruffert, M. 2005, A\&A, 444, L33

\bibitem[Domainko \& Ruffert(2008)]{domainko2008} Domainko, W. \& Ruffert, M. 2008, AdSpR, 41, 518

\bibitem[Domainko(2011)]{domainko2011} Domainko, W. F. 2011, A\&A, 533, L5

\bibitem[Eger et al.(2010)]{eger2010} Eger, P., Domainko, W. \& Clapson, A.-C. 2010. A\&A, 513, 66

\bibitem[Galante \& Horvath(2007)]{galante2007} Galante, D. \& Horvath, J. E. 2007, IJAsB, 6, 19

\bibitem[Gehrels et al.(2009)]{gehrels2009} Gehrels, N., Ramirez-Ruiz, E. \& Fox, D. B. 2009, ARA\&A, 47, 567

\bibitem[Grindlay et al.(2006)]{grindlay2006} Grindlay, J., Portegies Zwart, S. \& McMillan, S. 2006, NatPh, 2, 116

\bibitem[Guetta \& Stella(2009)]{guetta2009} Guetta, D. \& Stella, L. 2009, A\&A, 498, 329

\bibitem[Horvath(2003)]{horvath2003} Horvath, J. E. 2003, arXiv:astro-ph/0310034v1

\bibitem[Lee et al.(2010)]{lee2010} Lee, W. H., Ramirez-Ruiz, E. \& van de Ven, G. 2010, ApJ, 720, 953

\bibitem[Marshall(2006)]{marshall2006} Marshall, C.R. 2006, AREPS 34, 355

\bibitem[Melott et al.(2004)]{melott2004} Melott, A. L., Lieberman, B. S., Laird, C. M. et al. 2004, IJAsB, 3, 55

\bibitem[Pooley \& Hut(2006)]{pooley2006} Pooley, D. \& Hut, P. 2006, ApJ, 646, L143

\bibitem[Portgies Zwart \& McMillan(2000)]{portegies2000} Portegies Zwart, S. F. \& McMillan, S. L. W. 2000, ApJ, 528, L17

\bibitem[Racusin et al.(2011)]{racusin2011} Racusin, J. L., Oates, S. R., Schady, P. et al. 2011, ApJ, 738, 138

\bibitem[Raup \& Sepkoski(1982)]{raup1982} Raup, D. M. \& Sepkoski, J. J. 1982, Science, 215, 1501

\bibitem[Salvaterra et al.(2008)]{salvaterra2008} Salvaterra, R., Cerutti, A., Chincarini, G., Colpi, M., Guidorzi, C. \& Romano, P. 2008, MNRAS, 388, L6

\bibitem[Salvaterra et al.(2010)]{salvaterra2010} Salvaterra, R., Devecchi, B., Colpi, M. \& D'Avanzo, P. 2010, MNRAS, 406, 1248

\bibitem[Thomas et al.(2005)]{thomas2005} Thomas, B. C., Melott, A. L., Jackman, C. H., et al. 2005, ApJ, 634, 509

\bibitem[Thorsett(1995)]{thorsett1995} Thorsett, S. E. 1995, ApJ, 444, L35

\bibitem[Vande Putte \& Cropper(2009)]{vandeputte2009} Vande Putte, D. \& Cropper, M. 2009, MNRAS, 392, 113

\bibitem[Zhang et al.(2011)]{zhang2011} Zhang, B.-B., Zhang, B., Liang, E.-W., et al. 2011, ApJ, 730, 141	



\end{thebibliography}
\end{document}